\documentclass[conference]{IEEEtran}
\usepackage[letterpaper, left=0.625in, right=0.625in, bottom=0.9in, top=0.7in]{geometry} 

\usepackage{graphicx,cite}
\usepackage{algorithm}
\usepackage{algorithmic}
\usepackage{amsmath}
\usepackage{subfig}

\usepackage{amsfonts}
\usepackage{booktabs}

\usepackage{cite}
\begin{document}
\title{Node Centrality Metrics for Hotspots \\Analysis in Telecom Big Data}

\author{\IEEEauthorblockN{Emil Mededovic, Vaggelis G. Douros, Petri M\"{a}h\"{o}nen\\}
\IEEEauthorblockA{Institute for Networked Systems, RWTH Aachen University\\
Kackertstrasse 9, 52072 Aachen, 
Germany\\
E-mail: emil.mededovic@rwth-aachen.de, \{vaggelis.douros, pma\}@inets.rwth-aachen.de}}

\maketitle 

\begin{abstract}
In this work, we are interested in the applications of big data in the telecommunication domain, analysing two weeks of datasets provided by Telecom Italia for Milan and Trento. Our objective is to identify hotspots which are places with very high communication traffic relative to others and measure the interaction between them. We model the hotspots as nodes in a graph and then apply node centrality metrics that quantify the importance of each node. We review five node centrality metrics and show that they can be divided into two families: the first family is composed of closeness and betweenness centrality whereas the second family consists of degree, PageRank and eigenvector centrality. We then proceed with a statistical analysis in order to evaluate the consistency of the results over the two weeks. We find out that the ranking of the hotspots under the various centrality metrics remains practically the same with the time for both Milan and Trento. We further identify that the relative difference of the values of the metrics is smaller for PageRank centrality than for closeness centrality and this holds for both Milan and Trento. Finally, our analysis reveals that the variance of the results is significantly smaller for Trento than for~Milan.
\end{abstract}
\IEEEpeerreviewmaketitle

\section{Introduction and related work}
Nowadays, telecom companies use widely big data \mbox{in order} to mine the behaviour of their customers, improve the quality of service that they provide and reduce the customers' churn. Towards this direction, demographic statistics, network deployments and call detail records (CDRs) are key factors that need to be carefully integrated in order to make accurate predictions. Though there are various open source data for the first two factors, researchers rarely have access to traffic demand data, since it is a sensitive information for the operators. Therefore, researchers need to rely on synthetic models, which do not always capture accurately large-scale mobile \mbox{networks~\cite{Asssb}.}

For example, the authors in~\cite{BaseS} analyse an heterogeneous cellular network which consists of different types of nodes, such as macrocells and microcells. Nowadays a popular model is the one from Wyner~\cite{wyner}, but it fails to fully capture a real heterogeneous cellular network because it is simplistic. Another approach is to use the spatial Poisson point process model (SPPP)~\cite{SPPP}, which can be derived from the premise that all base stations are uniformly distributed. However, a city can be classified in different areas, which have different population densities. These different areas can be characterised as \mbox{dense urban,} urban and suburban. To be able to classify the heterogeneous networks into these areas, the authors introduce SPPP for homogeneous and inhomogeneous sets. They show that the SPPP-model captures accurately both urban and suburban areas, whereas this is not the case for dense urban areas, because of a considerable population concentrated in small areas.

\begin{figure*}
\centering
	\subfloat[Milan]
	{\includegraphics[width=0.45\linewidth]{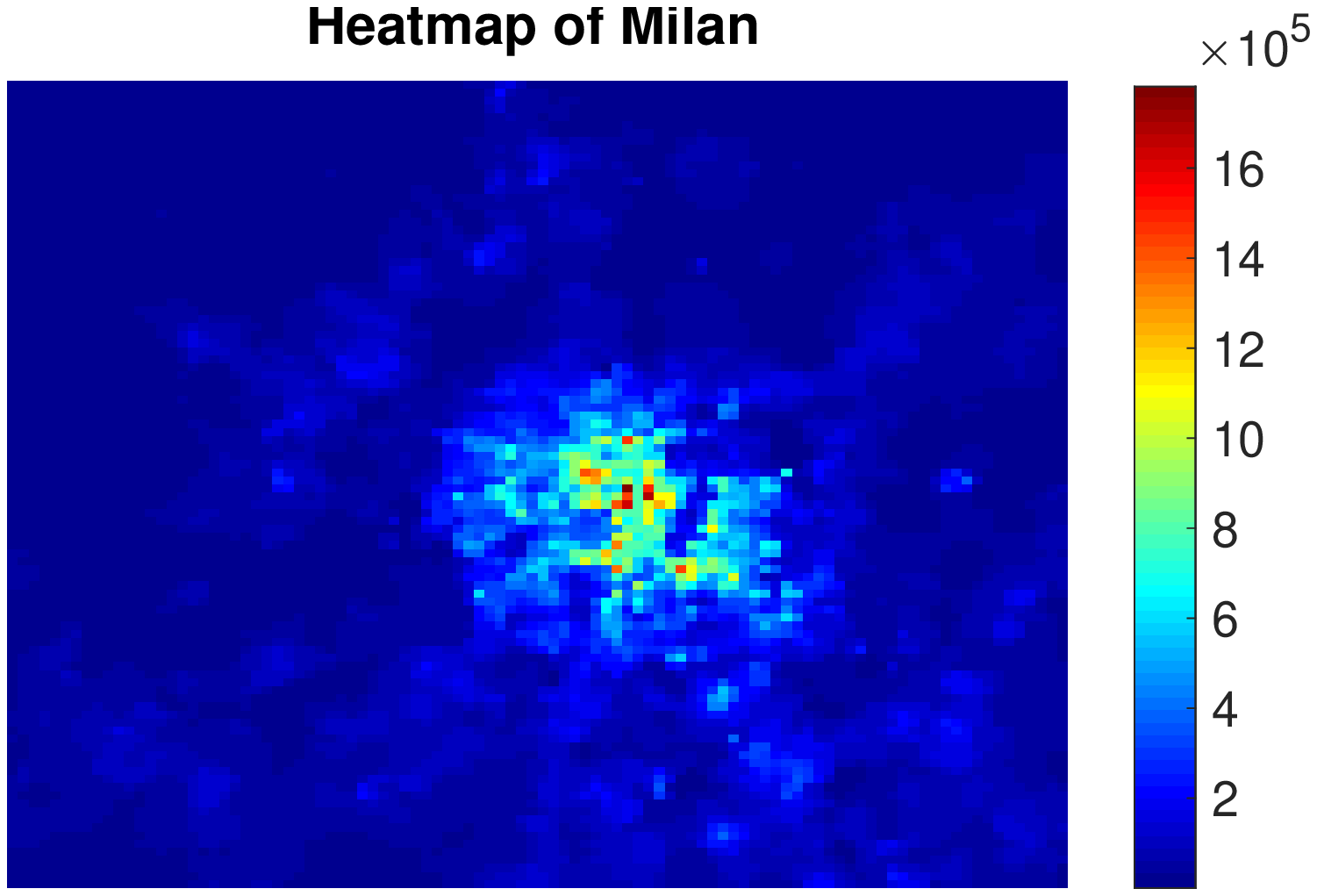}
	\label{fig:heatmapMilan}} \qquad
	\subfloat[Trento] 
	{\includegraphics[width=0.45\linewidth]{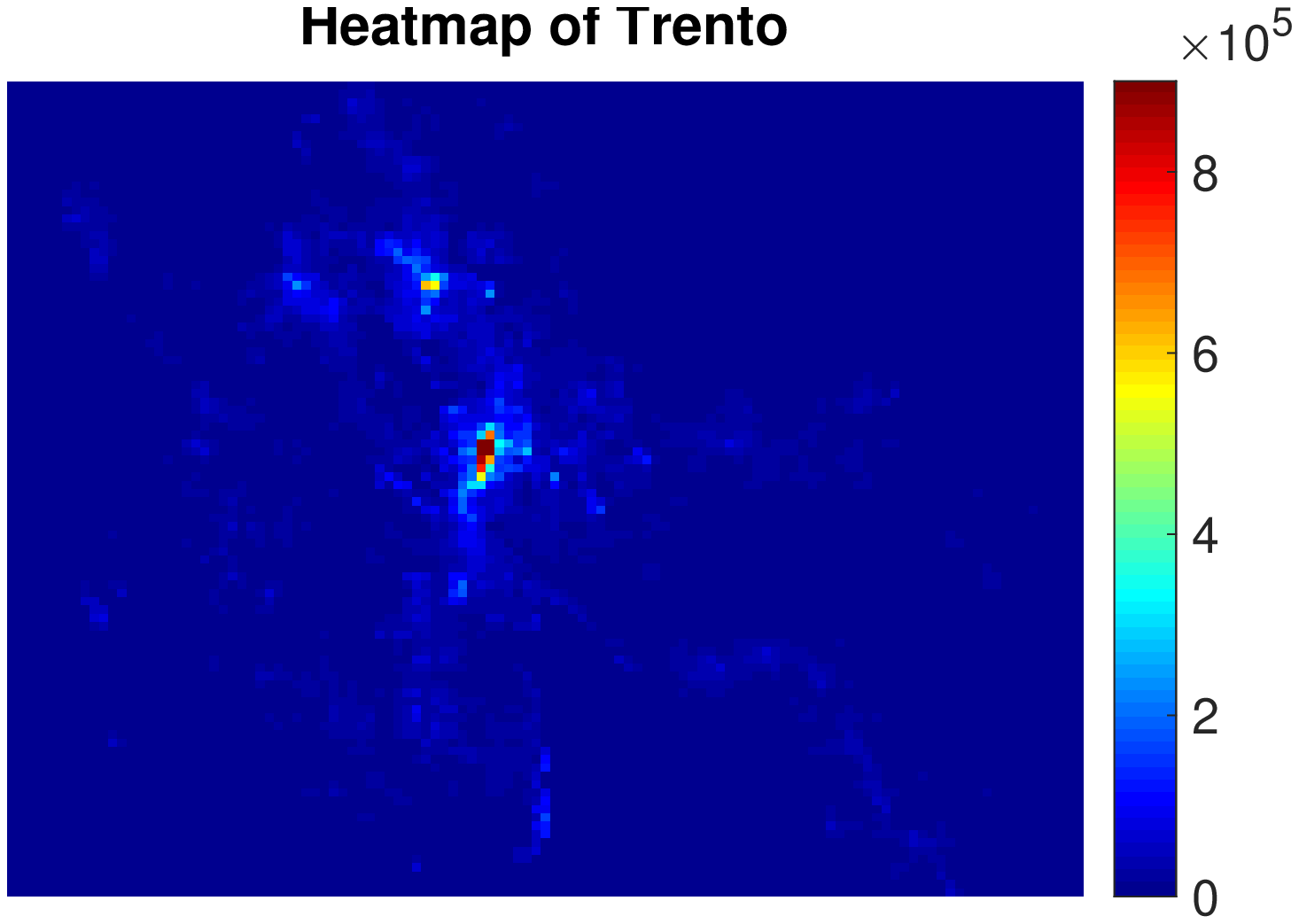}
	\label{fig:heatmapTrento}}\qquad
	\caption{These heatmaps visualise the communication intensity mapped over the different areas of Milan and Trento for week 1 (18.11.2013-24.11.2013). The intensity of the communication traffic increases as the colour changes from blue to red.}
	\label{fig:Heatmaps}
	\vspace{-0.1cm}
\end{figure*}

\begin{figure*} 
\centering
	\subfloat[closeness]
	{\includegraphics[width=0.45\linewidth]{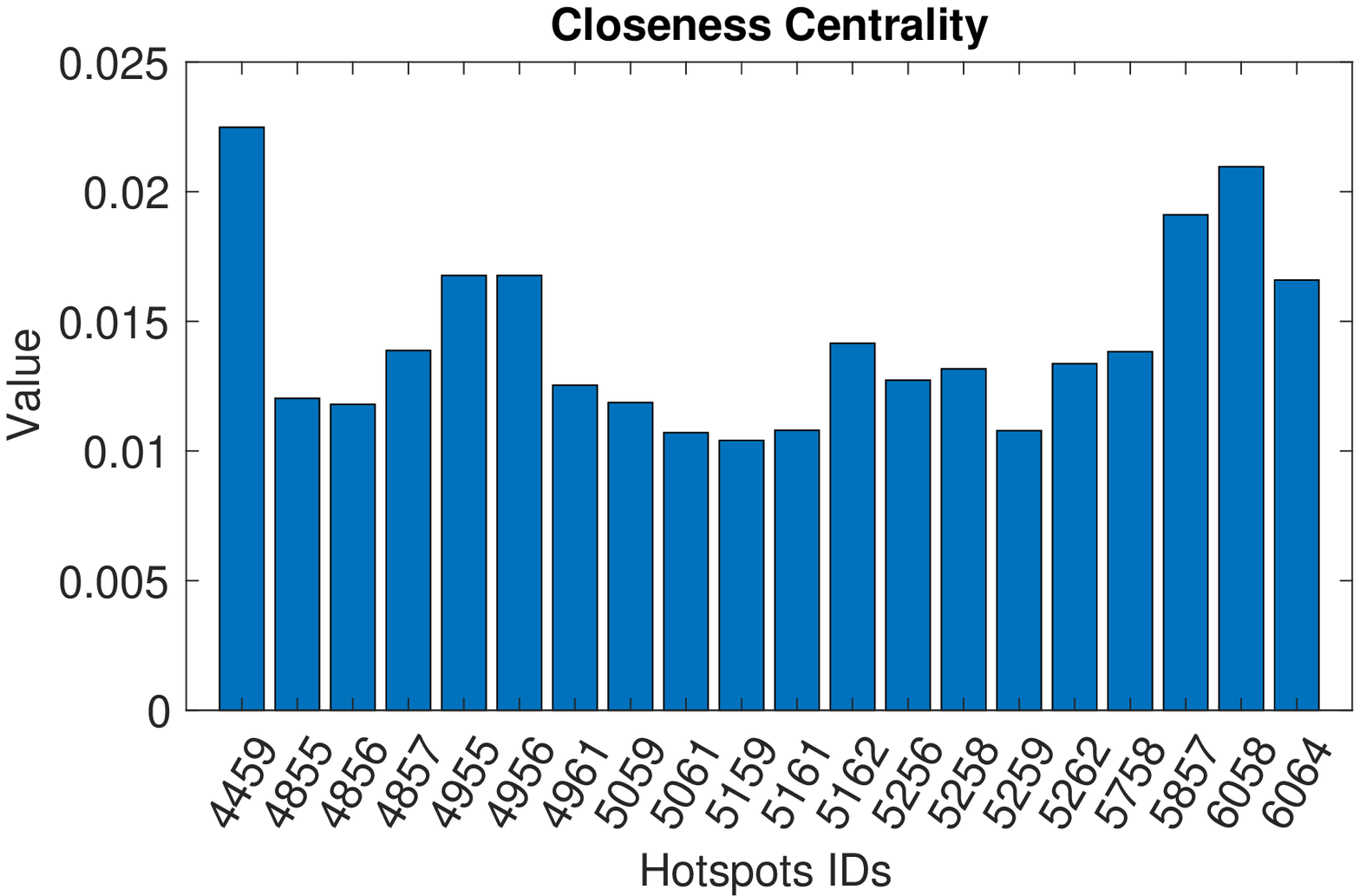}
	\label{fig:firstWeekClosenessMilan}} \qquad
	\subfloat[betweenness] 
	{\includegraphics[width=0.45\linewidth]{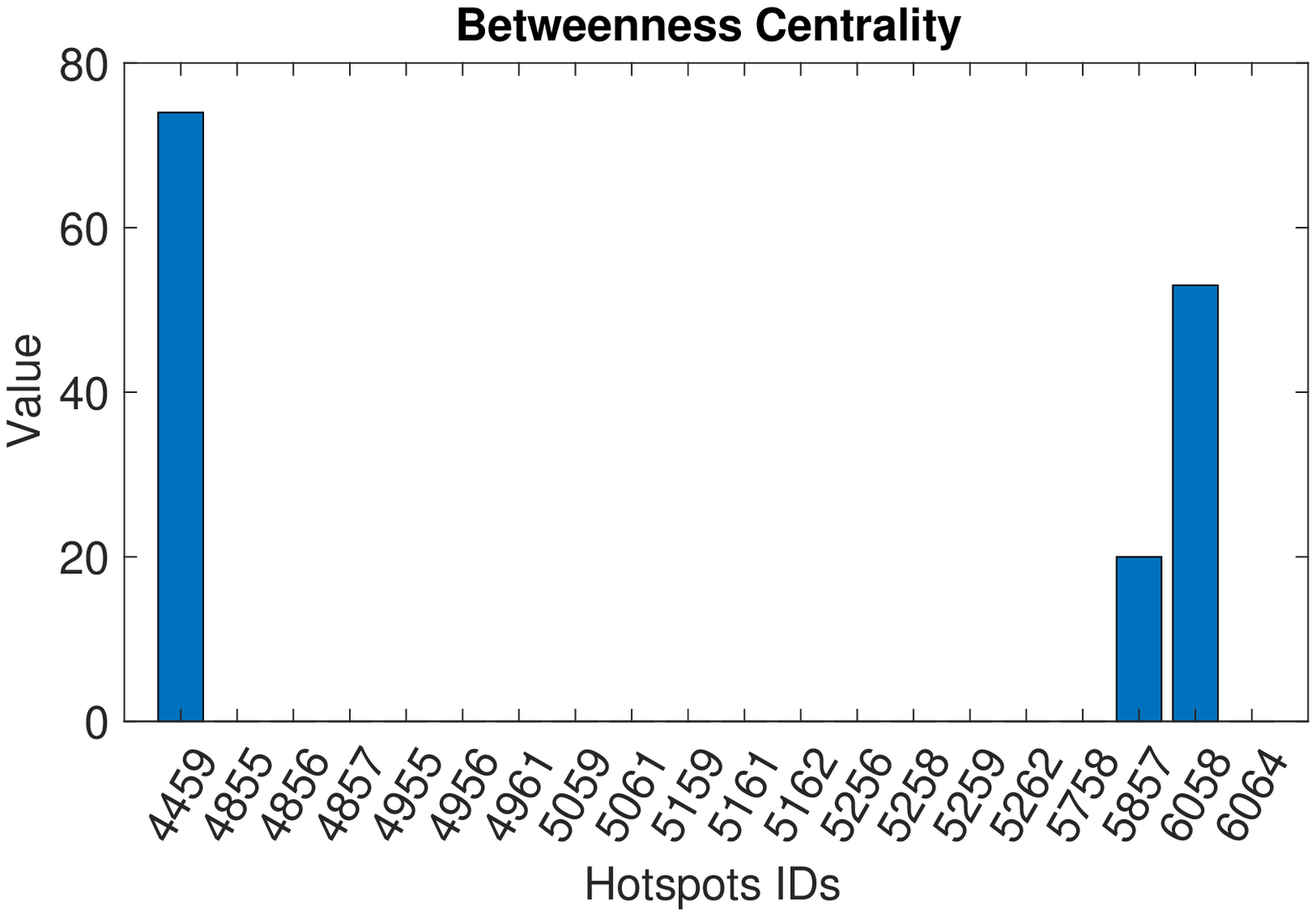}
	\label{fig:firstWeekBetweennessMilan}}\qquad
	\subfloat[degree]
	{\includegraphics[width=0.45\linewidth]{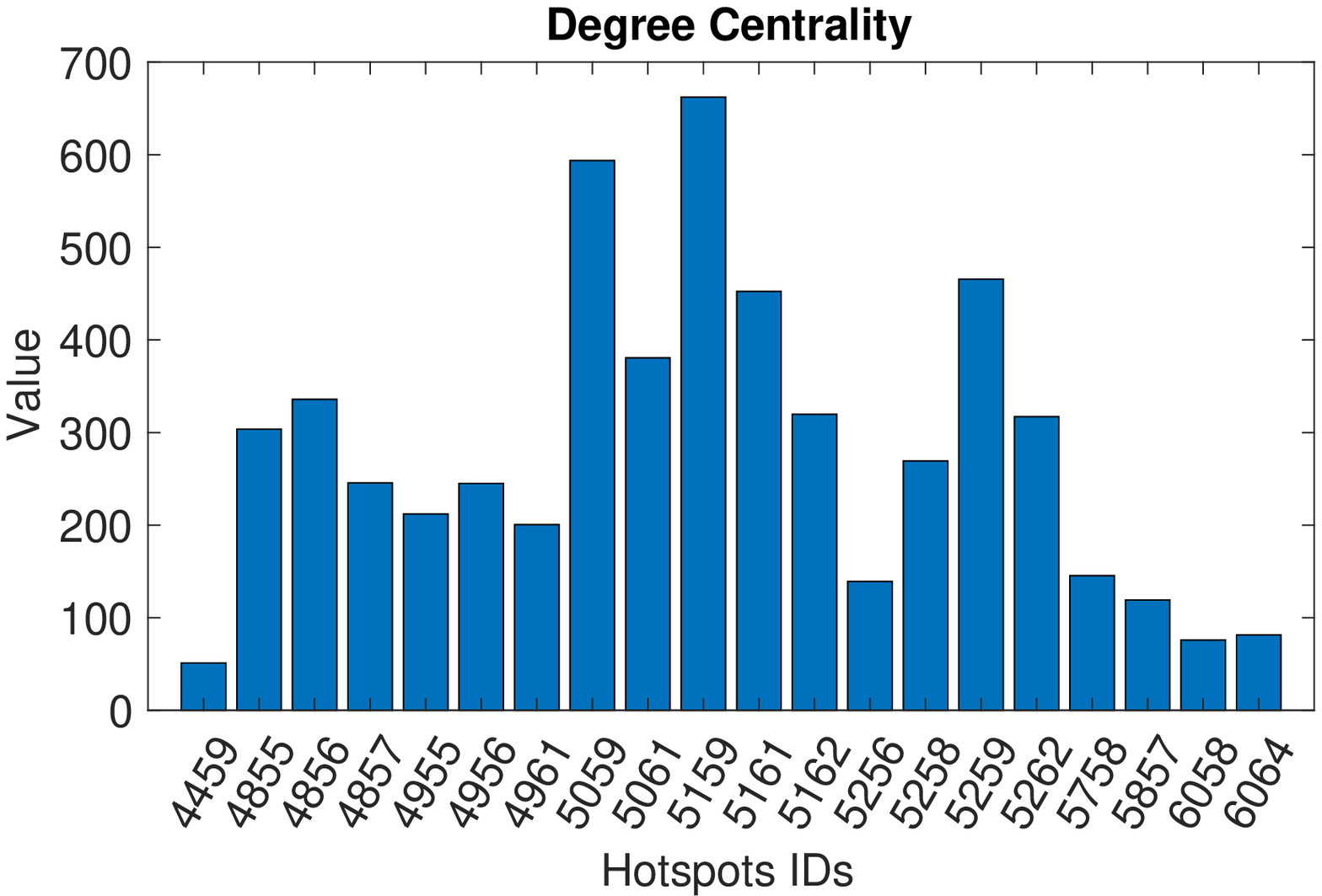}
	\label{fig:firstWeekDegreeMilan}} \qquad
	\subfloat[PageRank]
	{\includegraphics[width=0.45\linewidth]{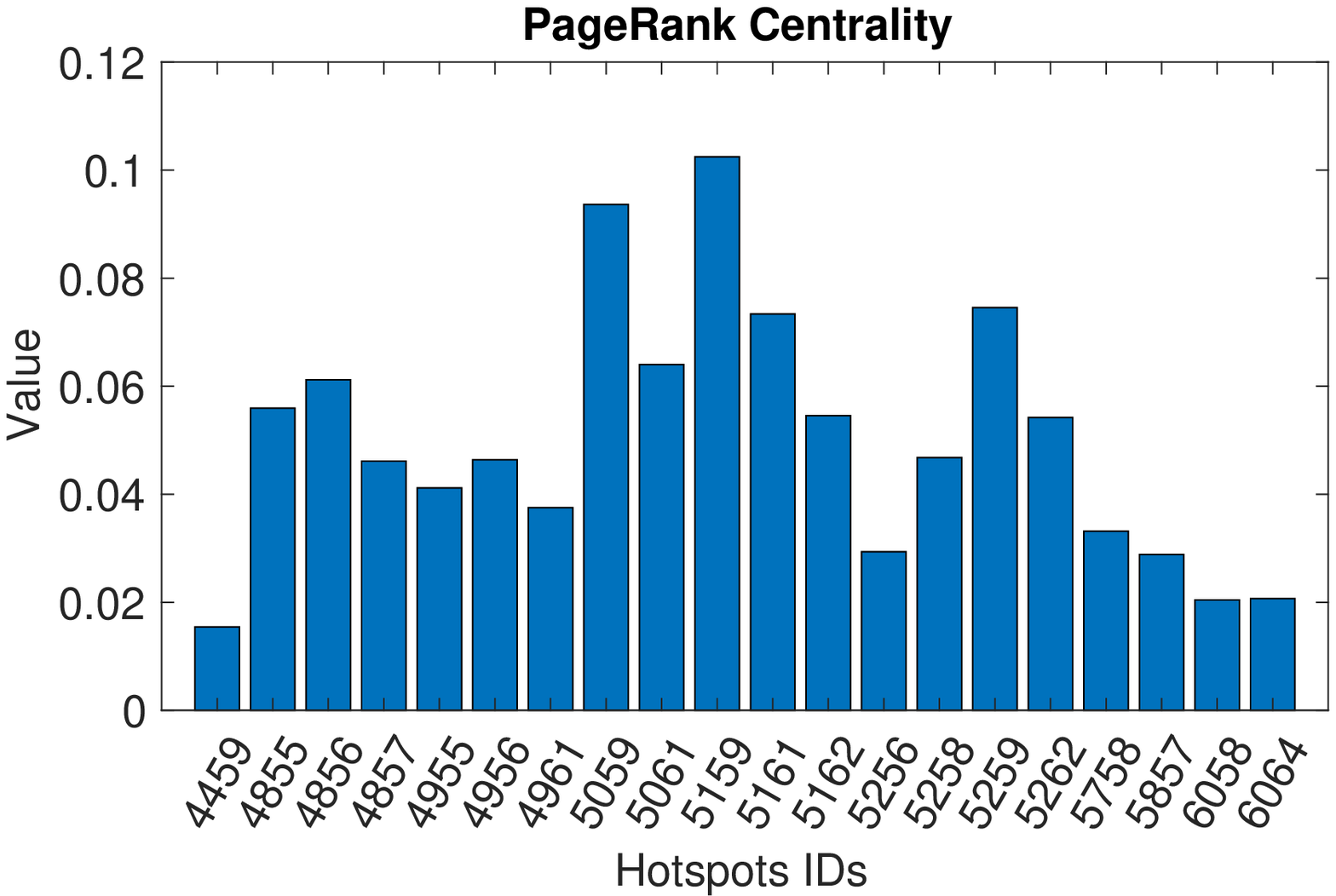}
	\label{fig:firstWeekPageRankMilan}} \qquad
	\subfloat[eigenvector]
	{\includegraphics[width=0.45\linewidth]{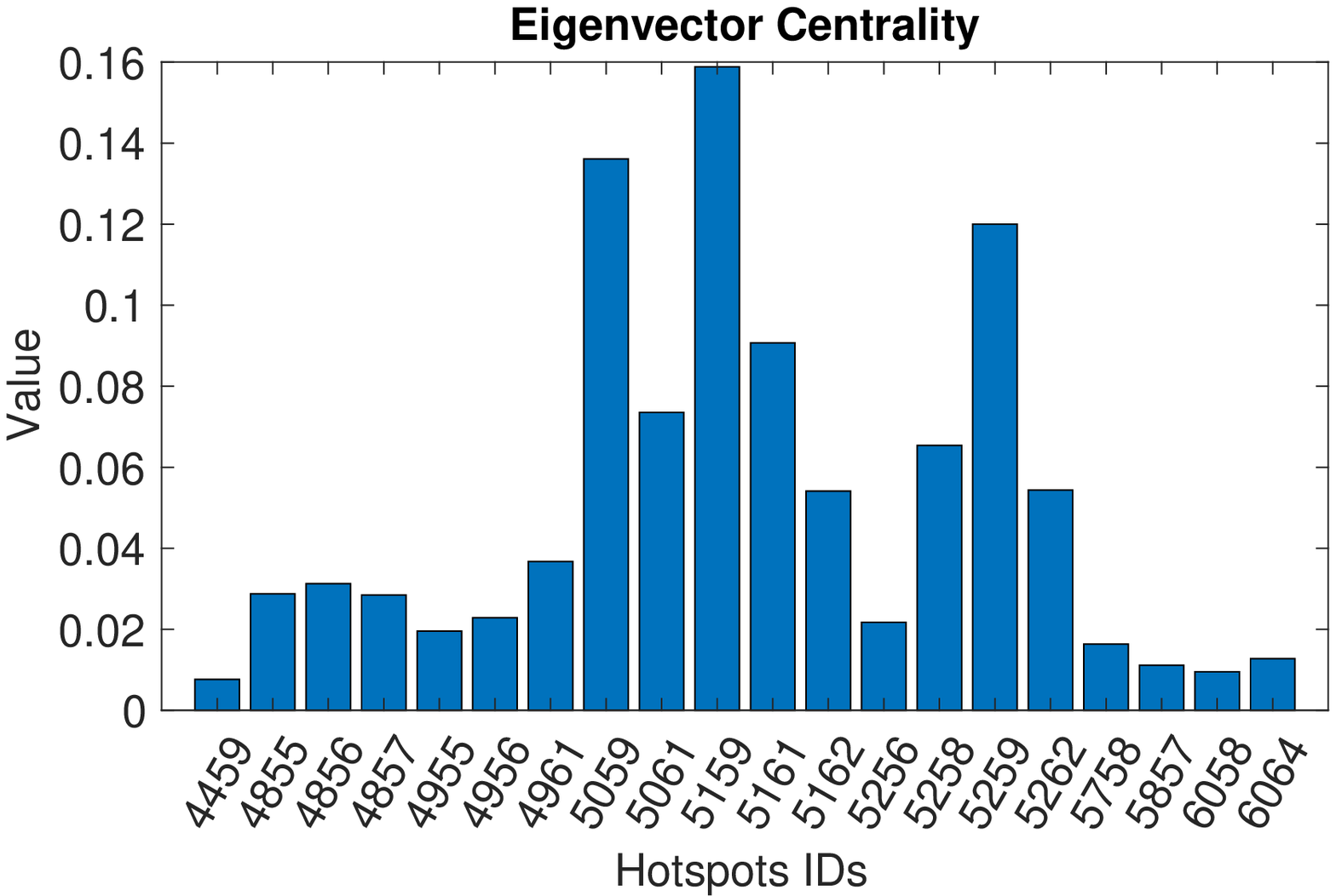}
	\label{fig:firstWeekEigenvectorMilan}} 
	\caption{Node centrality metrics for the different hotspots for week 1 (18.11.2013-24.11.2013) for Milan. The horizontal axis represents the id of the hotspots. The vertical axis represents the value of the corresponding centrality metric.}
	\label{fig:1weekMilan}
	\vspace{-0.1cm}
\end{figure*} 

\begin{figure*}
\centering
	\subfloat[closeness]
	{\includegraphics[width=0.45\linewidth]{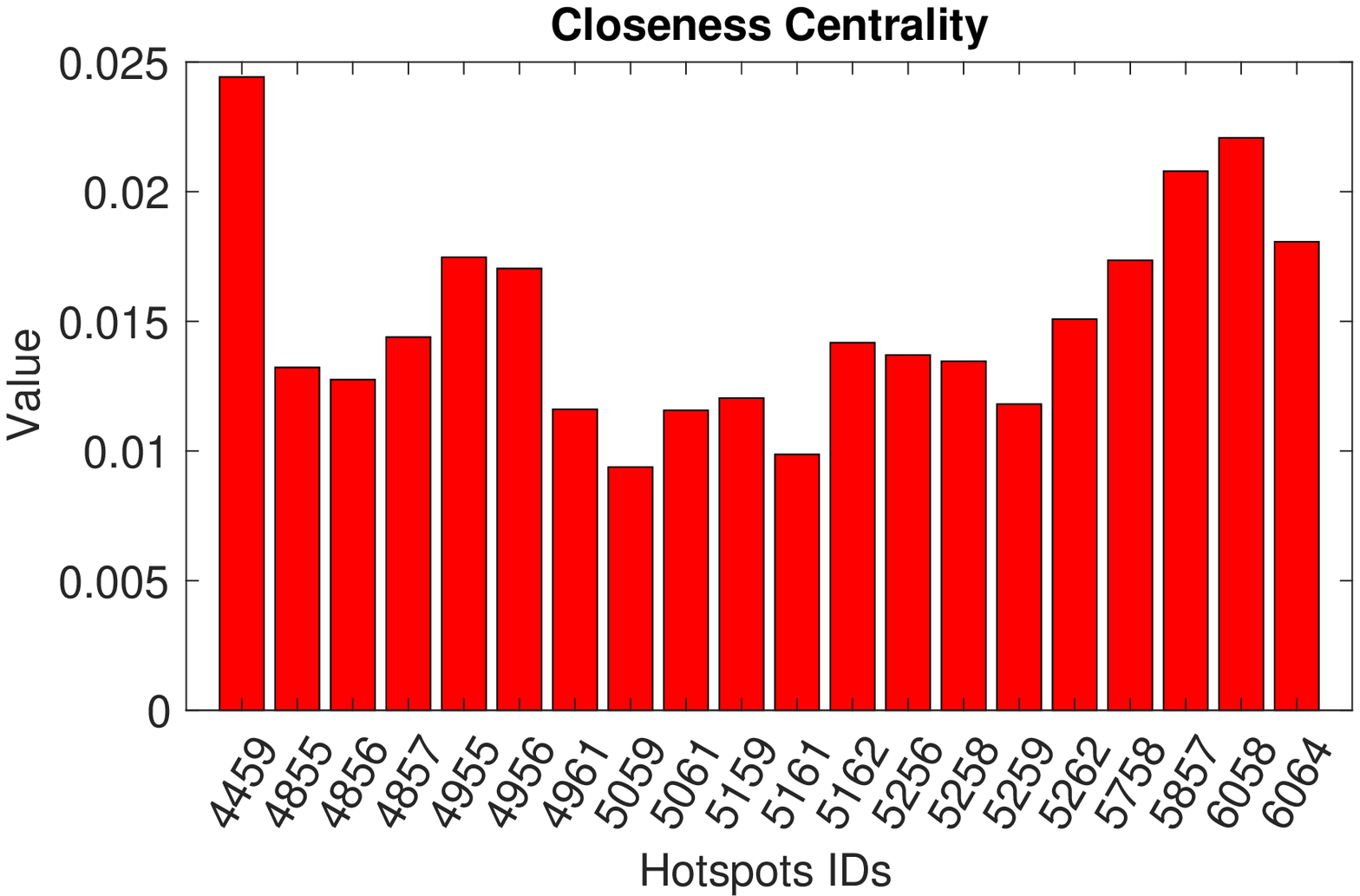}
	\label{fig:differenceClosenessValuesMilan}} \qquad
	\subfloat[PageRank] 
	{\includegraphics[width=0.45\linewidth]{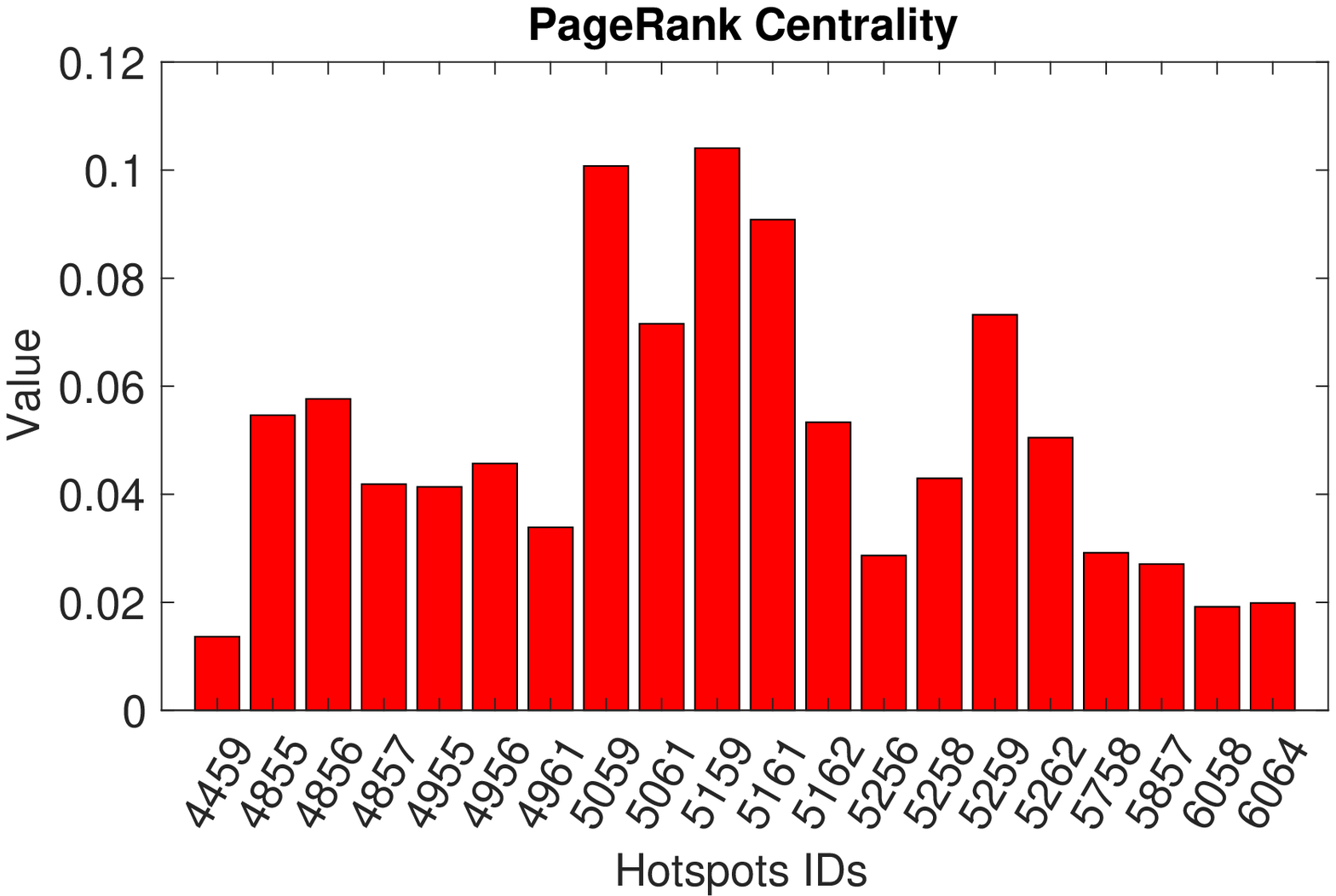}
	\label{fig:differencePageRankValuesMilan}}
	\caption{Closeness centrality and PageRank centrality for week 2 (8.12.2013-14.12.2013) for Milan. The horizontal axis represents the id of the hotspots. The vertical axis represents the values of the metrics.}
	\label{fig:diffMilan}
	\vspace{-0.1cm}
\end{figure*} 

\begin{figure*}
\centering
	\subfloat[closeness]
	{\includegraphics[width=0.45\linewidth]{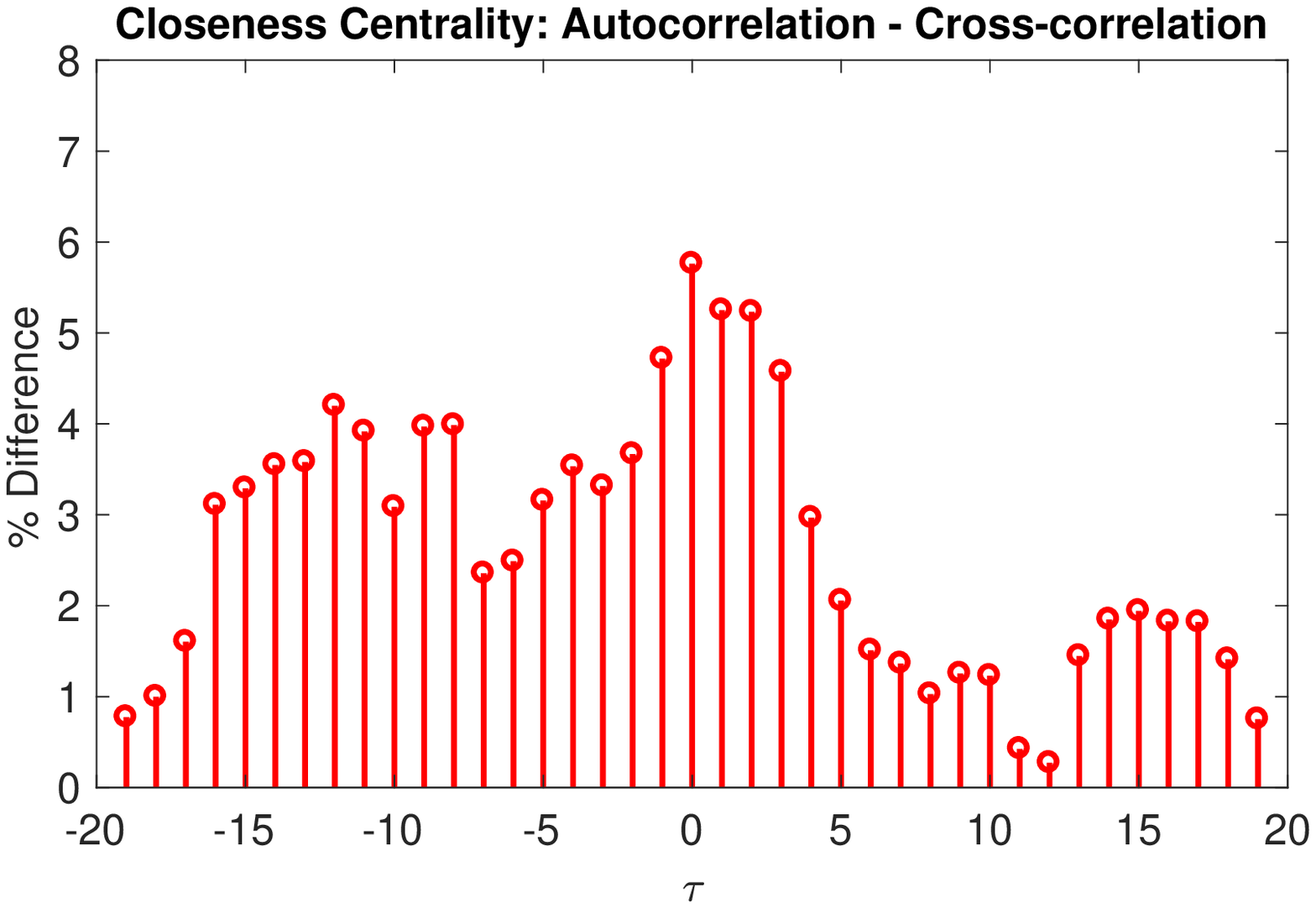}
	\label{fig:differenceAutoCrosscorrelationClosenessMilan}} \qquad
	\subfloat[PageRank] 
	{\includegraphics[width=0.45\linewidth]{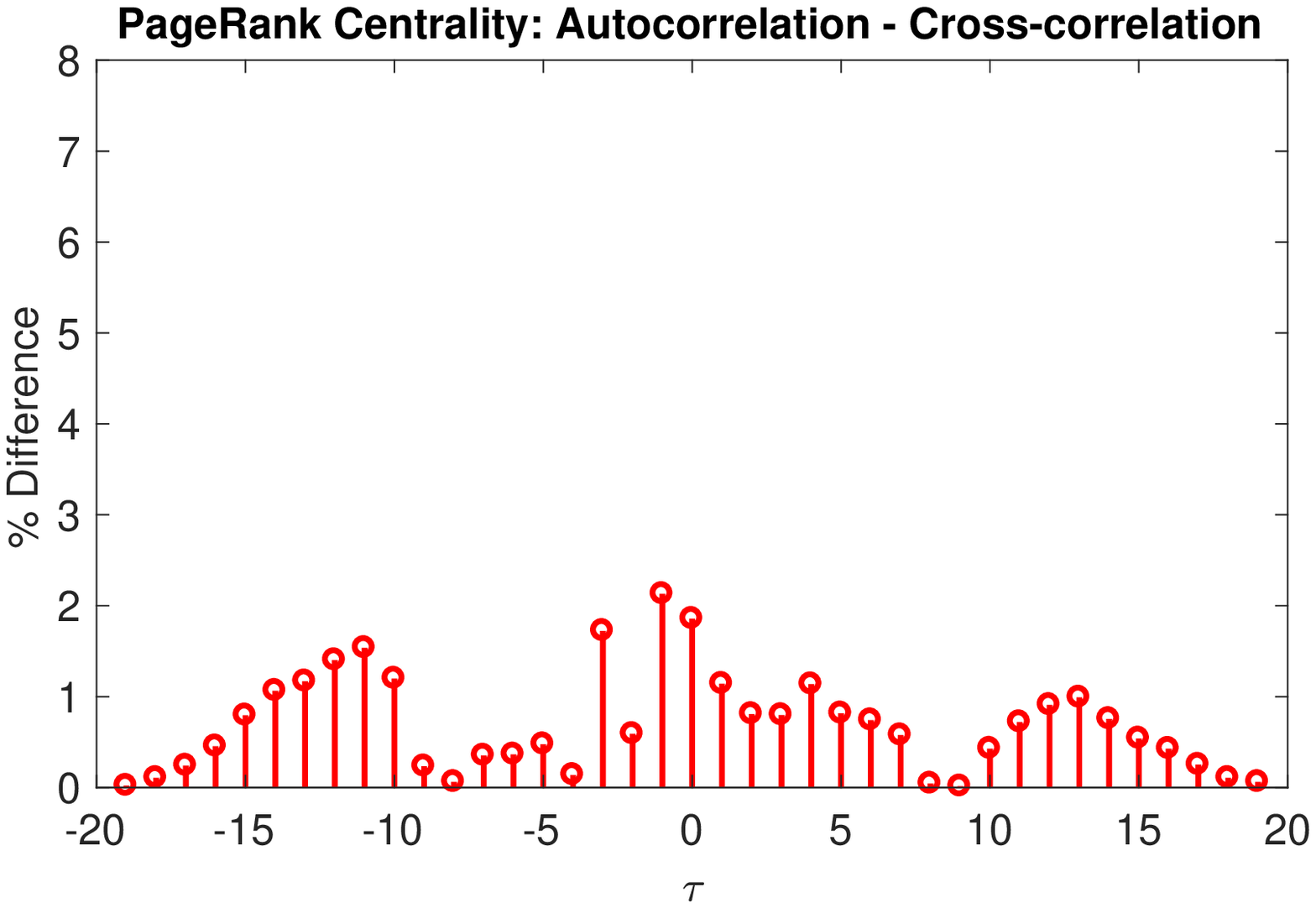}
	\label{fig:differenceAutoCrosscorrelationPageRankMilan}}
	\caption{Difference between autocorrelation and cross-correlation for the two representative centrality metrics for Milan. The horizontal axis represents the shift $\tau$. The vertical axis represents the difference in $\%$.}
	\label{fig:diffMilan2}
	\vspace{-0.1cm}
\end{figure*} 

\begin{figure*}
\centering
	\subfloat[closeness]
	{\includegraphics[width=0.45\linewidth]{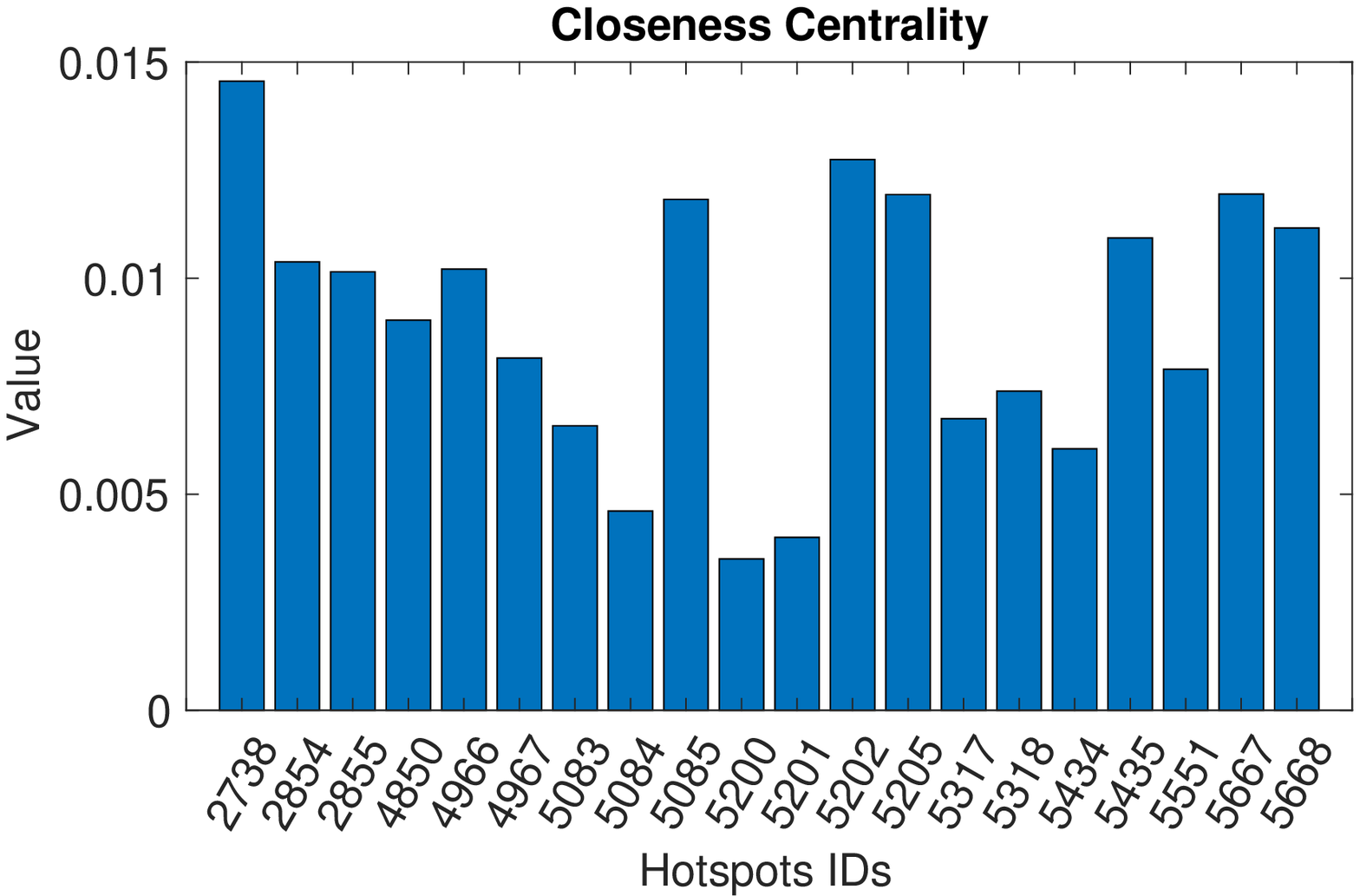}
	\label{fig:firstWeekClosenessTrento}} \qquad
	\subfloat[PageRank] 
	{\includegraphics[width=0.45\linewidth]{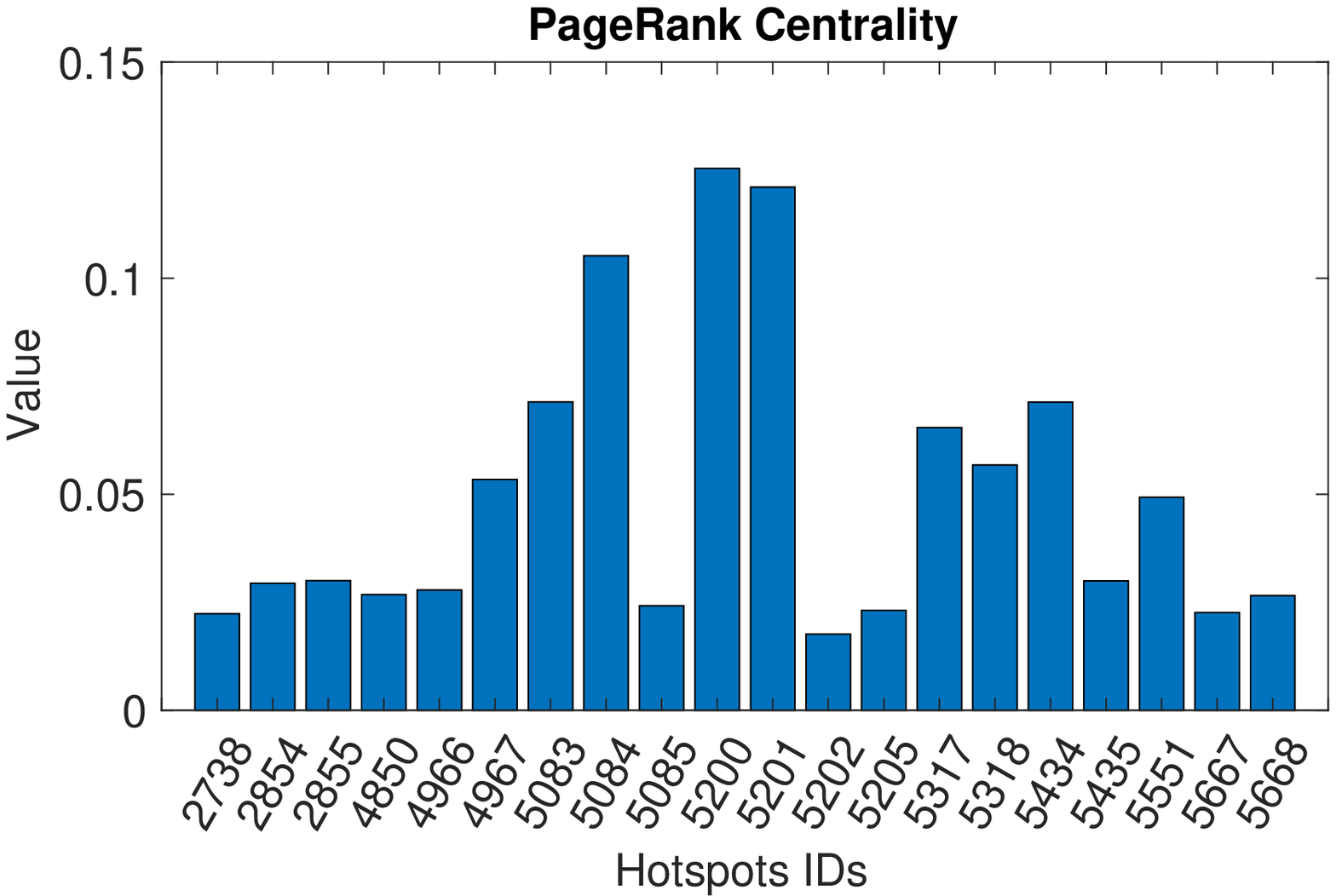}
	\label{fig:firstWeekPageRankTrento}} \qquad
	\subfloat[closeness]
	{\includegraphics[width=0.45\linewidth]{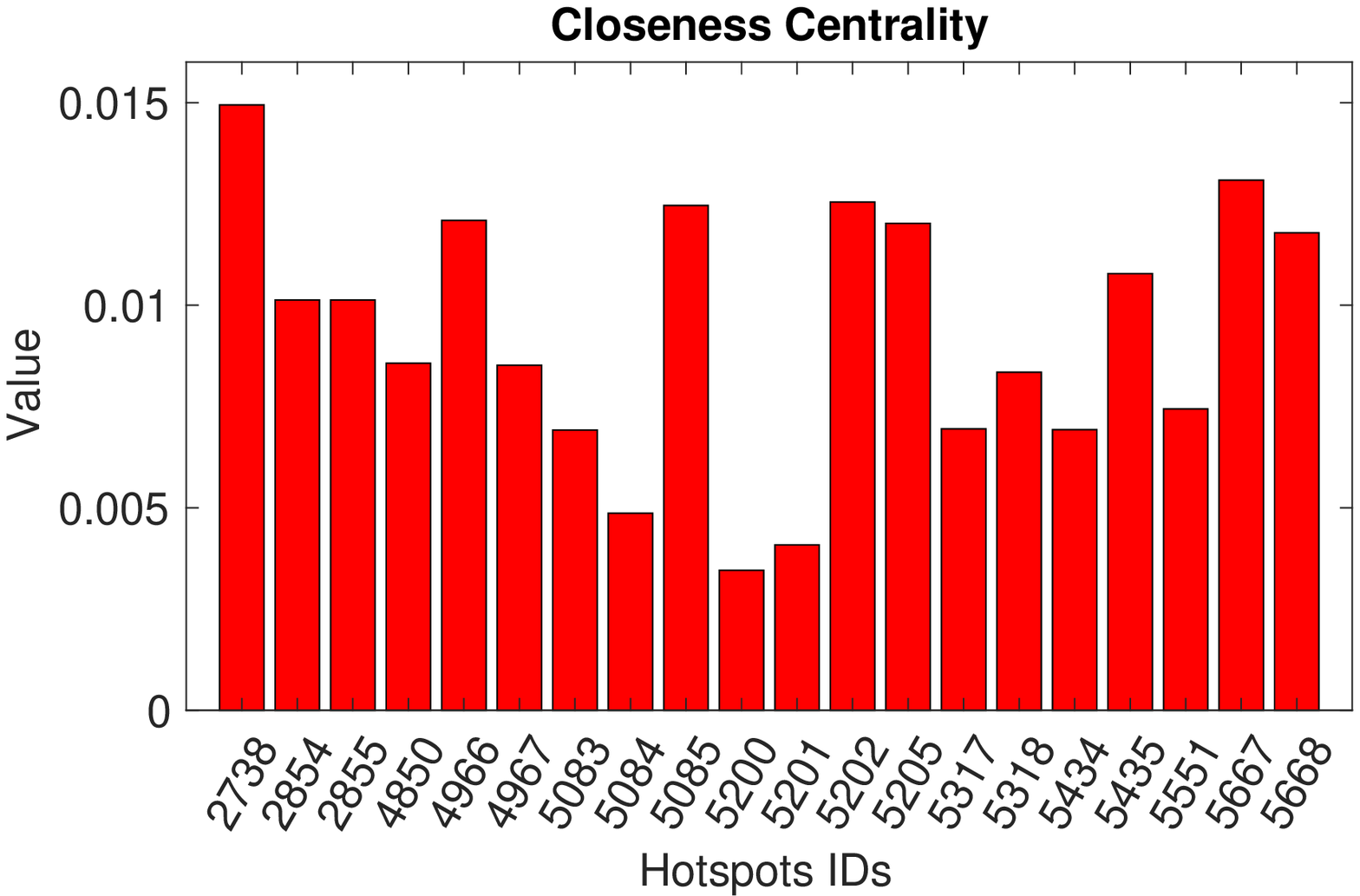}
	\label{fig:differenceClosenessValuesTrento}} \qquad
	\subfloat[PageRank] 
	{\includegraphics[width=0.45\linewidth]{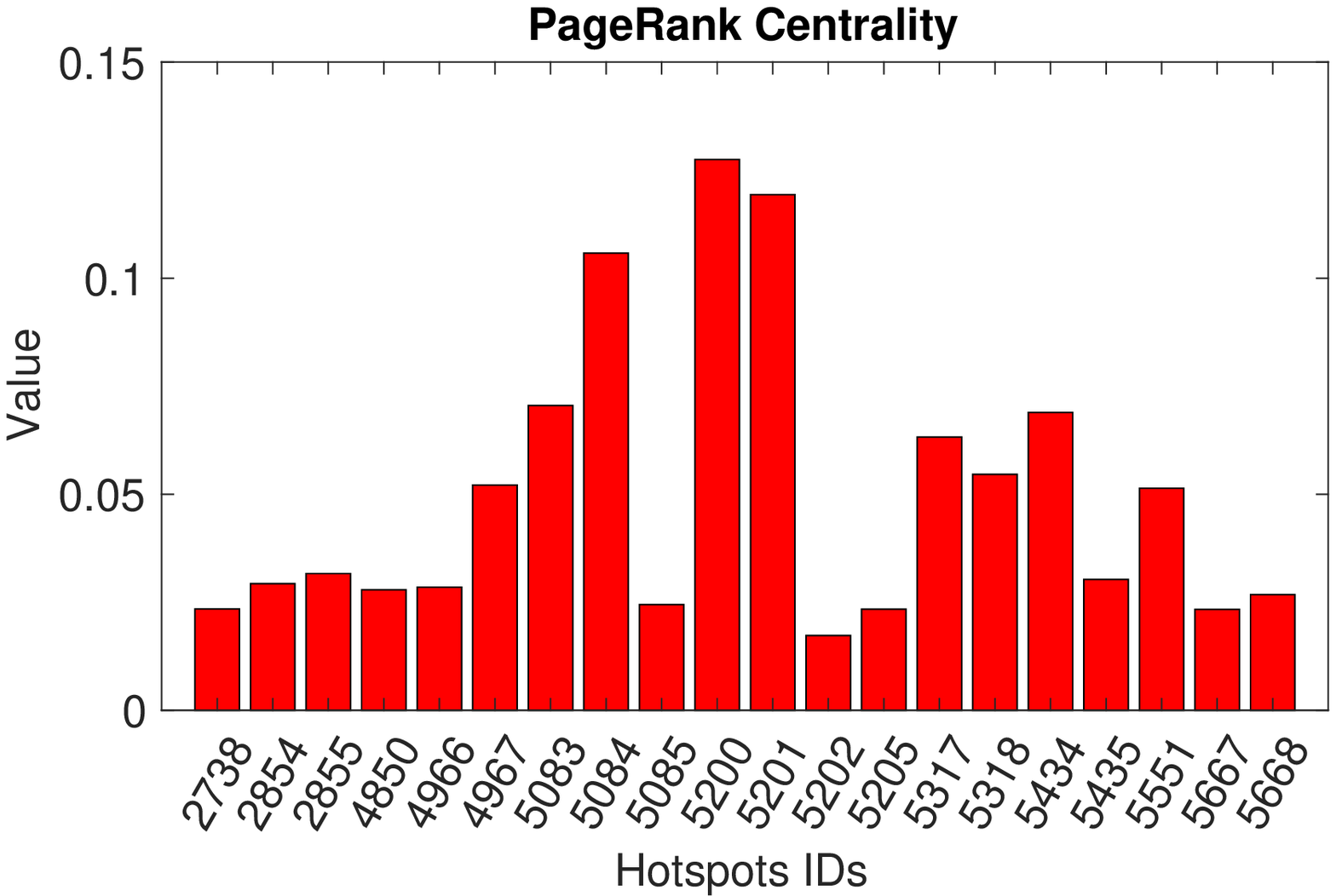}
	\label{fig:differencePageRankValuesTrento}} \qquad
	\subfloat[closeness]
	{\includegraphics[width=0.45\linewidth]{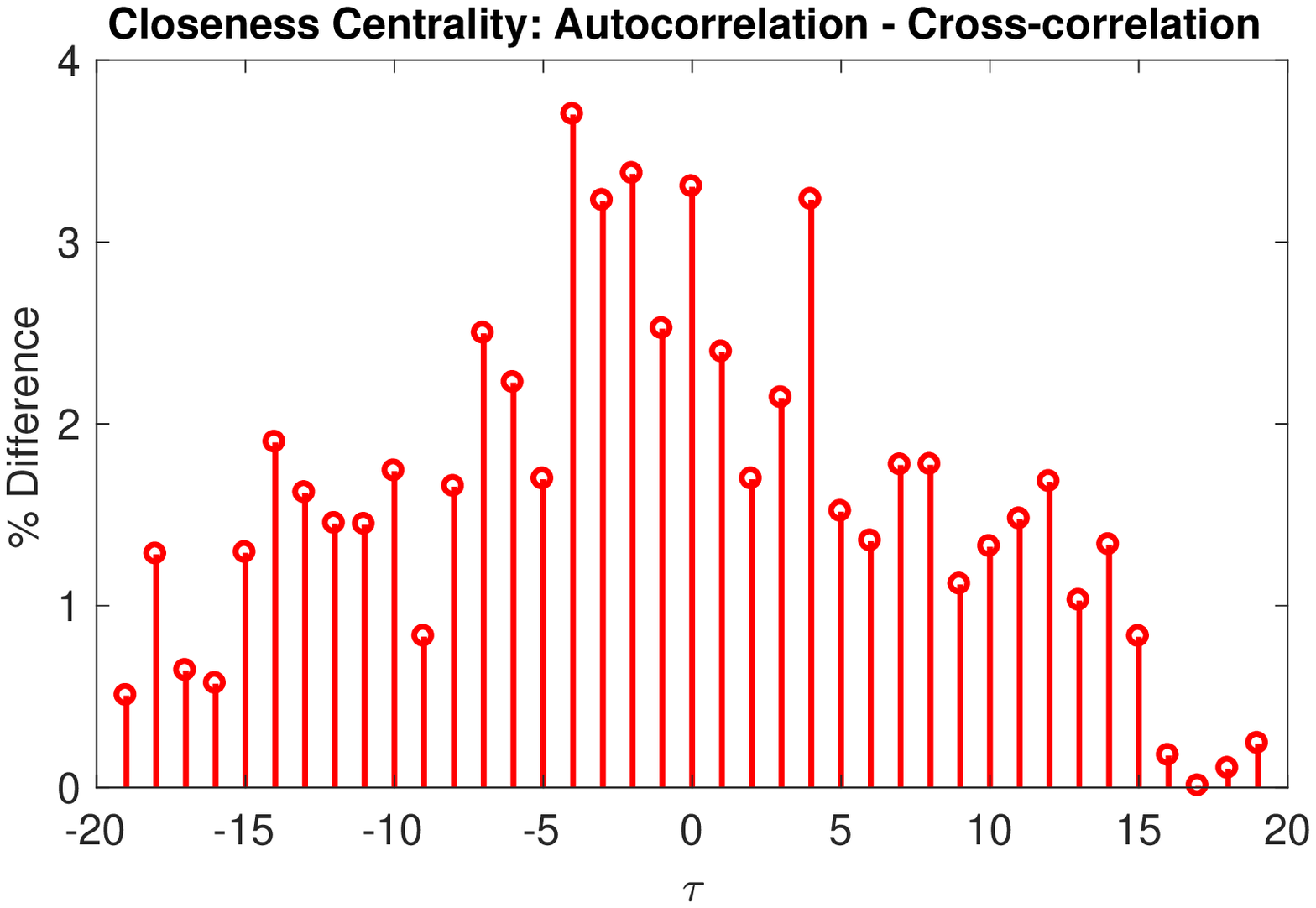}
	\label{fig:differenceAutoCrosscorrelationClosenessTrento}} \qquad
	\subfloat[PageRank] 
	{	\includegraphics[width=0.45\linewidth]{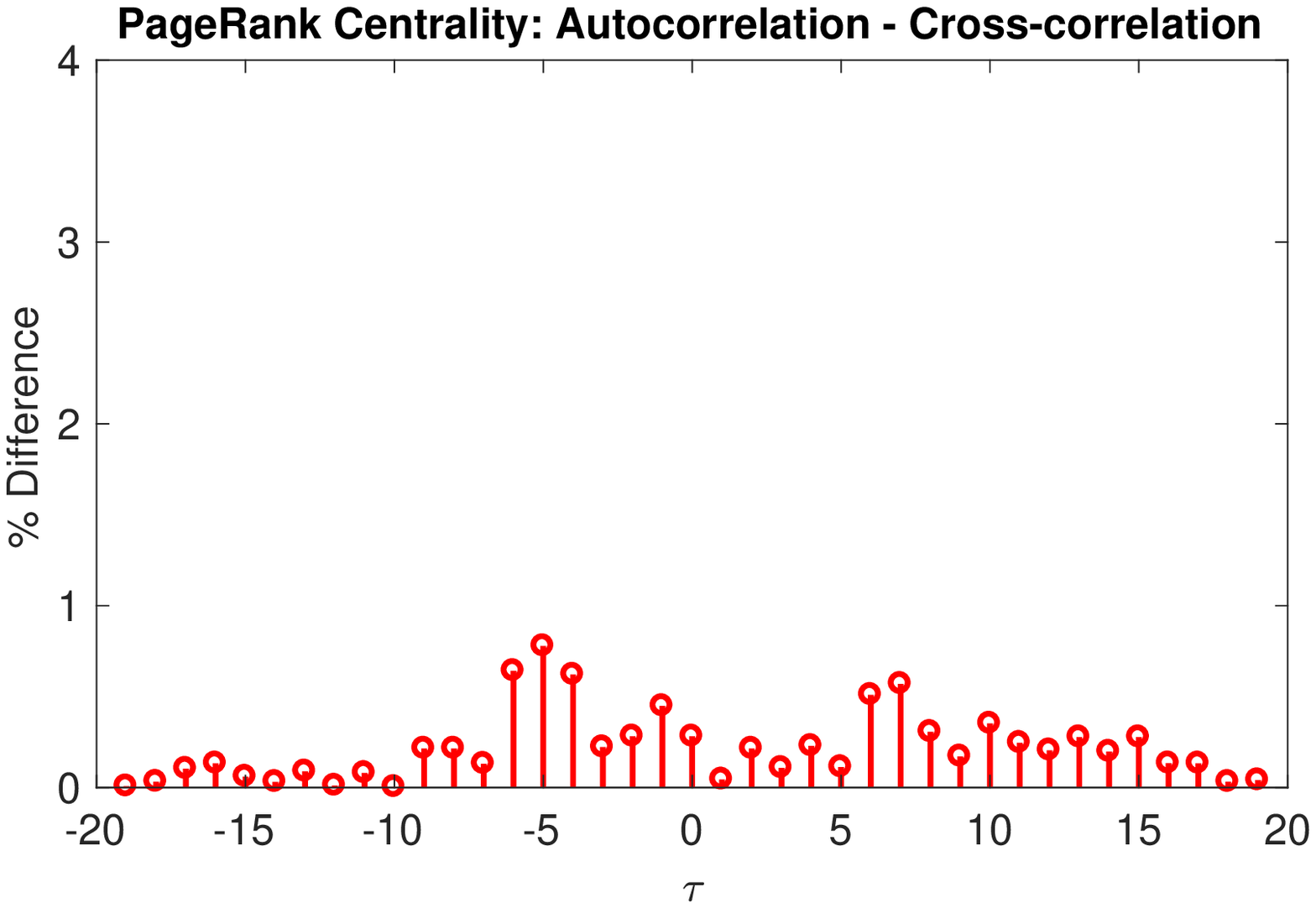}
	\label{fig:differenceAutoCrosscorrelationPageRankTrento}} 
	\caption{Results for Trento. Figures (a)-(d) show the results for closeness centrality and PageRank centrality for weeks 1 and~2. Figures (e)-(f) show the difference between autocorrelation and cross-correlation for each metric. The horizontal axis represents the shift $\tau$. The vertical axis represents the difference in $\%$.}
	\label{fig:analysisTrento}
	\vspace{-0.1cm}
\end{figure*}

One of the few exceptions of real deployment data being publicly available is the dataset published by Telecom Italia in 2014 as "the Big Data Challenge"~\cite{Data}. This includes data collected from November to December 2013 for Milan and Trento. Our goal is to use these datasets in order to identify hotspots (areas of high communication strength) and analyse their interactions. Before proceeding with our analysis, we review the state-of-the-art. 
%

The work closest to ours is~\cite{Happiness}. The authors aim at finding out whether areas with happy people communicate more often with other areas of the same kind or not. To investigate this topic, they model Milan as a graph where the nodes are the areas and the weights of the edges demonstrate the communication strength between one area and another. Then, in order to estimate the happiness level of the areas, they analyse the geolocalized tweets of the dataset. They rate the happiness of a tweet, by rating the happiness of each word of a tweet and computing the average of all ratings (given to each word) for one tweet. Using this approach, they are able to differentiate between happy and unhappy tweets. The conclusion is that there is an homophily pattern: happy urban areas tend to interact with other happy areas more than they interact with unhappy areas.  The same holds for unhappy areas. As in~\cite{Happiness}, we model the geographic area as a graph but we use network analysis in order to identify trends in the interactions between the areas. Moreover, we focus on hotspots, we analyse both Milan and Trento, and we use exclusively telecom data to measure the interactions between the areas.

Moreover, there have been published a number of machine learning approaches for the analysis of the Telecom Italia datasets. In~\cite{CDR}, the authors use CDRs of the datasets in order to detect anomalies in the network (i.e., unexpected and irregular behaviour of the users) and predict future traffic using machine learning algorithms. In particular, they use k-means clustering for the detection of the region for the anomalies. The cluster with the least points is considered as the region of anomalies. They then post-process the dataset to remove the abnormal activities of the users creating anomaly-free data and train a neural network based on them that makes accurate predictions. Finally, the authors use the ARIMA model to predict future traffic for a user. In~\cite{AI}, the authors review learning techniques for online network optimisation for the challenging problem of load balancing of the network, concluding that there is no clear winner. Finally, in~\cite{MDP}, the authors use unsupervised learning for the classification of mobile network usage profiles. 

In this paper, we use tools from network science~\cite{jackson2010social} to analyse two weeks of the Telecom Italia dataset for the cities of Milan and Trento aiming at identifying hotspots. Our contributions are three-fold: First, we show that the use of node centrality metrics~\cite{jackson2010social} is a simple but powerful tool in order to analyse hotspots. We apply five of the most popular node centrality metrics and show that they can be classified into two distinct groups: The first group is composed of closeness and betweenness centrality which favours hotspots with low weights whereas the second group is composed of degree, PageRank and eigenvector centrality which favours hotspots with high weights. Secondly, our analysis reveals that the ranking of the hotspots remains practically the same under the various centrality metrics as we move from one week to the other for both Milan and Trento. Moreover, the relative difference of the values of the metrics is smaller for PageRank centrality than for closeness centrality and this holds for both Milan and Trento. Finally, we find out that the variance of the results is significantly smaller for Trento than for Milan.  

 \section{Analysis}
In this section, we first briefly describe the datasets that we use from~\cite{Data}. The first dataset is called "telecommunication activity" and is composed of: square id which is the identification number of a given square of Milan/Trento grid, the approximative time of the event, incoming/outgoing amount of connections for SMS, incoming/outgoing amount of connections for call, internet traffic, and country code. This dataset is available for both Milan and Trento and will be used for the identification of the hotspots.

The second dataset is called "Milan/Trento to Milan/Trento calls" and is composed of the following fields:
\begin{itemize}
	\item Square id$_1$: identification number of the square of Milan/Trento grid that represents the origin of the interaction.
	\item Square id$_2$: identification number of the square of Milan or Trento grid that represents the destination of the interaction.
	\item The approximative time of the event.
	\item Directional interaction strength: value representing the directional interaction strength between Square id$_1$ and Square id$_2$.
\end{itemize}

This dataset will be used for the application of node centrality metrics in order to measure the communication strength since it gives information about the source and the destination of the communication whereas the first dataset only considers \mbox{one area.} 

The last dataset that we consider is called "Milan grid" and is composed of:
\begin{itemize}
	\item square id: identification string of a given square of the Milan or Trento grid.
	\item The cell geometry expressed as geoJSON and projected in WGS84 (EPSG:4326).
\end{itemize}
This dataset is also available for both cities and will be used to visualise the hotspots. 

Having completed the discussion of the datasets, we then present our approach for the identification of the hotspots. Through the "telecommunication activity" dataset, we identify hotspots as areas with very high communication traffic. This includes the aggregated amount of connections for SMS, call and internet data. In order to quantitively determine the areas with enough high communication traffic to be considered as hotspots, we have to define a threshold. This threshold represents the minimum amount of communication traffic for an area to be considered as a hotspot and can be tuned by changing a parameter. All these requirements culminates to the following definition. We define area $i$ as a hotspot if it fulfils this inequality:

\begin{equation}
I_i \geq \frac{1}{N} \cdot \sum_{j=1}^{N}I_j+\bigtriangleup,	 \\
\end{equation}
where $I_i$ is the amount of communication for area \emph{i}. The value of $\bigtriangleup$ can be calculated as follows: 
\begin{equation}
\bigtriangleup = \left(\mathrm{MaxTraffic} - \frac{1}{N} \cdot\sum_{j=1}^{N}I_j\right)\cdot P,
\end{equation}
where \emph{MaxTraffic} is the maximum amount of communication in all areas, and $P$ is a parameter to determine the cutoff threshold.

In Figs. \ref{fig:heatmapMilan} and \ref{fig:heatmapTrento}, we show an indicative visualisation of the communication traffic for the areas of Milan and Trento. We observe that the communication traffic is distributed more evenly for Trento than for Milan: the communication traffic of Milan shows a high concentration around the centre of Milan.

After extracting the hotspots, we use the following node centrality metrics~\cite{jackson2010social} that quantify the importance of each hotspot: degree centrality, closeness centrality, betweenness centrality, PageRank centrality, and eigenvector centrality.

Next, we discuss the main idea of these metrics. Let $M(x)$ be the set of all nodes except  node $x$, and $d(\cdot,\cdot)$ be the function for the distance between two connected nodes in a graph. 
For closeness centrality, we always consider the shortest distance between two nodes. Then, the closeness centrality for a node $x$ is defined as follows:
\begin{equation*}
 C(x) = \frac{1}{\sum_{y \in M(x)} d(x,y)}.	 
\end{equation*}
 
Let $\sigma_{st}(x)$ be the number of the shortest paths from node $s$ to $t$ which goes through node $x$, and $\sigma_{st}$ be the number of the shortest paths from $s$ to $t$. Then, the betweenness centrality for a node $x$ is defined as follows: 
\begin{equation*}
B(x) = \sum_{s\neq t\neq x} \frac{\sigma_{st}(x)}{\sigma_{st}}. 
\end{equation*}

Let N($\cdot$) be the set of neighbours for a node. The degree centrality for a node $x$ is defined as follows:
\begin{equation*}
D(x) = \sum_{y \in N(x)} d(x,y).
\end{equation*}

Let $\mathrm{PR}(\cdot)$ be the PageRank score for a node, $q$ be the damping factor, $n$ be the number of nodes, and $L(\cdot)$ be the number of neighbours of a node. For the PageRank centrality, we always consider the direct distance between two neighbours without going through intermediate nodes. Then, the PageRank centrality for a node $x$ is defined as follows:
\begin{equation*}
\mathrm{PR(x)}= \dfrac{1-q}{n}+q\cdot\sum_{y\in N(x)} \dfrac{\mathrm{PR(y)}}{L(y)}. 
\end{equation*}

Finally, let $\lambda$ be the biggest eigenvalue of the adjacency matrix of the corresponding graph. Then, the eigenvector centrality for a node $x$ is defined as follows:
\begin{equation*}
\mathrm{Eig(x)}= \dfrac{1}{\lambda}\cdot\sum_{y \in N(x)} d(x,y). 
\end{equation*}
 
In order to measure the importance of the hotspots, we calculate for each hotspot the different centrality metrics. We analyse the datasets for the same two weeks for both Milan and Trento so as to compare the results both over the same city and between cities. For these computations, we use the dataset "Milan/Trento to Milan/Trento calls" because we need a source and a destination of these communication traffics in order to compute a graph and apply these metrics. Moreover, to compare effectively the results of two different weeks, we use the cross-correlation function to detect if there is a correlation between the results from the two time series. As we deal with a discrete number of hotspots, we use the discrete version of the cross-correlation function, which is defined as follows~\cite{box2015time}:
\begin{equation*}
(f \ast g)[n]=\sum_{m=-\infty}^{+\infty} f[m] \cdot g[m+n].
\end{equation*}
%
%
Finally, to measure the robustness of the results of the \mbox{cross-correlation} function, we compute the corresponding autocorrelation function that expresses the correlation of a time series with a delayed copy of itself~\cite{box2015time}:
 \begin{equation*}
 (f \ast f)[n]=\sum_{m=-\infty}^{+\infty} f[m] \cdot f[m+n].
 \end{equation*}

\section{Results for Milan}
In this section, we present the results for Milan. We start our big data analysis with the week from 18.11.2013 till 24.11.2013 (denoted as week~1) where we identify the top-20 hotspots that generate the highest traffic. Therefore, we use (1) and (2) and set experimentally the parameter $P=0.75$ in order to get exactly 20 hotspots. In Figs. \ref{fig:firstWeekClosenessMilan} and \ref{fig:firstWeekBetweennessMilan}, we observe that the hotspots with ids 4459 and 6058 have the highest values for both closeness and betweenness centrality. On the other hand, we see from Figs.~\ref{fig:firstWeekDegreeMilan}, \ref{fig:firstWeekPageRankMilan} and \ref{fig:firstWeekEigenvectorMilan} that the following three hotspots (ids: 5059, 5159 and 5259) have the highest centrality values under degree, PageRank and eigenvector. Furthermore, based on the adjacent ids, we find out that these three hotspots are neighbours to each other and are situated in the centre of Milan (which was expected from the structure of the heatmap of Fig.~\ref{fig:heatmapMilan}). 

A general remark is that the node centrality metrics can be classified into two groups. The first group composed of closeness and betweenness centrality favours hotspots with low weights because the shortest path is a key component of both centrality metrics. The second group composed of degree, PageRank and eigenvector centrality favours hotspots with high weights because higher weights around a node is an indication of a high centrality value for this node. This can be also noticed from Fig.~\ref{fig:1weekMilan} because areas with high values for centrality metrics from one group have low scores for the centrality metric of the other group. For instance, the hotspot with the id 5159 has the lowest centrality score for the first group but the highest value for the second. This observation simplifies the analysis because we can consider one metric per family.  From now on, we focus on closeness and PageRank centrality  respectively. 

In order to evaluate the robustness of the results (i.e., whether the trends for the hotspots are consistent across the dataset), we continue with the analysis of the dataset for week~2 (8.12.2013-14.12.2013), using the same 20 hotspots with week~1.  By comparing the values of the centrality metrics for week~1 and week~2, we notice from Fig.~3 that the ranking of the hotspots for week~2 remain practically the same with week~1 for both metrics. The relative difference with respect to week~1 is less than $10\%$ for each hotspot for closeness centrality and less than $8\%$ for PageRank centrality (with the exception of one outlier for each metric). Finally, in Figs.~\ref{fig:differenceAutoCrosscorrelationClosenessMilan} and \ref{fig:differenceAutoCrosscorrelationPageRankMilan}, we present the relative difference between cross-correlation and autocorrelation to quantify the difference between the observed level of similarity and the perfect one. We note that the difference is less than $6 \%$ for closeness centrality and less than $2\%$ for PageRank. The above results indicate the consistency of the results since we find out that both the ranking of the hotspots and the relative difference of metrics per hotspot do not vary significantly.   

%

\section{Results for Trento}
We then proceed with the analysis for Trento. We identify again the top-20 hotspots for the same two weeks that we used for Milan. In Figs.~\ref{fig:firstWeekClosenessTrento} and \ref{fig:firstWeekPageRankTrento}, we present the results for week~1. We note that the hotspots with ids 2738 and 5202 have the highest closeness centrality values whereas the hotspots with ids 5200 and 5201 have the highest PageRank centrality values. We observe again that hotspots with high closeness centrality have low PageRank centrality and vice versa. 

In Figs.~\ref{fig:differenceClosenessValuesTrento} and \ref{fig:differencePageRankValuesTrento}, we present the centrality metrics for week~2. The ranking of the hotspots remains consistent: the most important hotspots for week~2 are similar to the ones for week~1. When it comes to the relative difference of the centrality metrics with respect to week~1, we notice that, with the exception of one outlier, the relative difference is less than $8\%$ for the closeness centrality. For PageRank, the difference is always less than $2\%$. Finally, from Figs.~\ref{fig:differenceAutoCrosscorrelationClosenessTrento} and \ref{fig:differenceAutoCrosscorrelationPageRankTrento}, we notice that the difference between cross-correlation and autocorrelation is always less than $4\%$ for closeness centrality and under $1\%$ for PageRank centrality. 




\section{Conclusions} 
In this work, we used tools from network science to analyse two weeks of telecom big data for the cities of Milan and Trento with the view to identifying hotspots. A general conclusion from our study is that the use of node centrality metrics is a simple but powerful tool in order to analyse hotspots. Another key conclusion is that the node centrality metrics can be classified into two distinct groups: the first group is composed of closeness and betweenness centrality which favours hotspots with low weights whereas the second group is composed of degree, PageRank and eigenvector centrality which favours hotspots with high weights. Our big data analysis has shown that the ranking of the hotspots remains practically the same under the various centrality metrics as we move from one week to the other for both Milan and Trento. Moreover, we found out that the relative difference of the values of the metrics is smaller for PageRank centrality than for closeness centrality and this holds for both Milan and Trento. Finally, we found out that the variance of the results is significantly smaller for Trento than for Milan.  

As a future work, we are interested in analysing the whole dataset and examine whether the conclusions from the two weeks can be generalised.  Towards this direction, a natural extension is the combination of our approach with machine learning methods for traffic forecasting. Another direction is to analyse the dataset of geolocalized tweets as in~\cite{Happiness} and compare the hotspots. Finally, it is interesting to apply the node centrality metrics to different cities with similar geographic and demographic features with Milan and Trento and evaluate the impact of these factors.  
%
%

\bibliographystyle{IEEEtran}
\bibliography{IEEEabrv,references}

\begin{thebibliography}{10}
\providecommand{\url}[1]{#1}
\csname url@samestyle\endcsname
\providecommand{\newblock}{\relax}
\providecommand{\bibinfo}[2]{#2}
\providecommand{\BIBentrySTDinterwordspacing}{\spaceskip=0pt\relax}
\providecommand{\BIBentryALTinterwordstretchfactor}{4}
\providecommand{\BIBentryALTinterwordspacing}{\spaceskip=\fontdimen2\font plus
\BIBentryALTinterwordstretchfactor\fontdimen3\font minus
  \fontdimen4\font\relax}
\providecommand{\BIBforeignlanguage}[2]{{%
\expandafter\ifx\csname l@#1\endcsname\relax
\typeout{** WARNING: IEEEtran.bst: No hyphenation pattern has been}%
\typeout{** loaded for the language `#1'. Using the pattern for}%
\typeout{** the default language instead.}%
\else
\language=\csname l@#1\endcsname
\fi
#2}}
\providecommand{\BIBdecl}{\relax}
\BIBdecl

\bibitem{Asssb}
P.~D. Francesco, F.~Malandrino, and L.~A. DaSilva, ``Assembling and using a
  cellular dataset for mobile network analysis and planning,'' \emph{IEEE
  Transactions on Big Data}, 2018.

\bibitem{BaseS}
J.~Zhang, W.~Wang, X.~Zhang, Y.~Huang, Z.~Su, and Z.~Liu, ``Base stations from
  current mobile cellular networks: Measurement, spatial modeling and
  analysis,'' \emph{IEEE Wireless Communications and Networking Conference
  Workshops (WCNCW)}, 2013.

\bibitem{wyner}
J.~Xu, J.~Zhang, and J.~G. Andrews, ``On the accuracy of the {W}yner model in
  cellular networks,'' \emph{IEEE Transactions on Wireless Communications},
  2011.

\bibitem{SPPP}
A.~Baddeley, I.~Bárány, and R.~Schneider, \emph{Stochastic Geometry}.\hskip
  1em plus 0.5em minus 0.4em\relax Springer, 2004.

\bibitem{Data}
G.~Barlacchi, M.~D. Nadai, R.~Larcher, A.~Casella, C.~Chitic, G.~Torrisi,
  F.~Antonelli, A.~Vespignani, A.~Pentland, and B.~Lepri, ``A multi-source
  dataset of urban life in the city of {M}ilan and the province of
  {T}rentino,'' \emph{Nature Scientific Data}, 2015.

\bibitem{Happiness}
A.~Alshamsi, E.~Awad, M.~Almehreyi, V.~Babushkin, P.-J. Chang, Z.~Shoroye,
  A.-P. Toth, and I.~Rahwan, ``Misery loves company: happiness and
  communication in the city,'' \emph{EPJ Data Science}, 2015.

\bibitem{CDR}
K.~Sultan, H.~Ali, and Z.~Zhang, ``Call detail records driven anomaly detection
  and traffic prediction in mobile cellular networks,'' \emph{IEEE Access},
  2018.

\bibitem{AI}
L.~Vigneri, N.~Liakopoulos, G.~S. Paschos, S.~Vassilaras, A.~Destounis,
  T.~Spyropoulos, and M.~Debbah, ``Model-driven artificial intelligence for
  online network optimization,'' \emph{arXiv preprint arXiv:1805.12090}, 2018.

\bibitem{MDP}
A.~Furno, D.~Naboulsi, R.~Stanica, and M.~Fiore, ``Mobile demand profiling for
  cellular cognitive networking,'' \emph{IEEE Transactions on Mobile
  Computing}, 2016.

\bibitem{jackson2010social}
M.~O. Jackson, \emph{Social and economic networks}.\hskip 1em plus 0.5em minus
  0.4em\relax Princeton University Press, 2010.

\bibitem{box2015time}
G.~E. Box, G.~M. Jenkins, G.~C. Reinsel, and G.~M. Ljung, \emph{Time series
  analysis: forecasting and control}.\hskip 1em plus 0.5em minus 0.4em\relax
  John Wiley \& Sons, 2015.

\end{thebibliography}
\end{document}